\documentclass[12pt]{article}
\usepackage{amsmath}
\usepackage{amssymb}
\usepackage{epsfig}
\usepackage{cite}
\usepackage{color,colordvi}
\newcommand{\eq}{\begin{equation}}
\newcommand{\eqx}{\end{equation}}
\newcommand{\eqn}{\begin{eqnarray}}
\newcommand{\bi}{\begin{itemize}}
\newcommand{\eqnx}{\end{eqnarray}}
\newcommand{\ei}{\end{itemize}}

\newcounter{hran}

\def\MSbar{\relax\ifmmode\overline{\rm MS}\else{$\overline{\rm MS}${ }}\fi}

\begin{document}

\begin{center}

{\Large\bf  Black Hole Masses are Quantized}
%\vspace{2cm}

\vspace{0.4 cm}

\end{center}

\begin{center}

%\hfill CERN-PH-TH/2010-082\\[0pt]
%\vskip -.1 cm \hfill LMU-ASC 26/10 \vskip -.1 cm \hfill MPP-2010-49\\[0pt]

%{\Large \textbf{Probing Quantum Geometry at LHC}}%\vspace{2cm}

%\vspace{1cm}

\textbf{Gia Dvali}$^{a,b,d,c}$, \textbf{Cesar Gomez}$^{a,e}$ and \textbf{Slava
Mukhanov}$^{a,b}$

\vspace{.4truecm}

\vspace{.2truecm}

\emph{$^a$Arnold Sommerfeld Center for Theoretical Physics\\[0pt]
Department f\"ur Physik, Ludwig-Maximilians-Universit\"at M\"unchen\\[0pt]
Theresienstr.~37, 80333 M\"unchen, Germany}

%\vspace{.2truecm}

\emph{$^b$Max-Planck-Institut f\"ur Physik\\[0pt]
F\"ohringer Ring 6, 80805 M\"unchen, Germany}

%\vspace{.6truecm}

\emph{$^c$CERN, Theory Division\\[0pt]
1211 Geneva 23, Switzerland}

%\vspace{.2truecm}

\emph{$^d$CCPP, Department of Physics, New York University\\[0pt]
4 Washington Place, New York, NY 10003, USA}

\vspace{.2truecm}

\emph{$^e$ Instituto de F\'{\i}sica Te\'orica UAM-CSIC, C-XVI \\[0pt]
Universidad Aut\'onoma de Madrid, Cantoblanco, 28049 Madrid, Spain}\\[0pt]
%\end{center}

%{\bf Gia,  Cesar, Slava,  the working file!!}

\end{center}

%\end{center}

\vskip .1in

\centerline{\bf Abstract}
\vskip .1in
%\no
 
 We give a simple argument  showing that in any sensible quantum field theory
 the masses of black holes cannot assume continuous values and must be quantized.  
 Our proof solely relies on Poincare-invariance of the asymptotic background, and is  insensitive 
 to geometric characteristics of black holes or other peculiarities  of the short distance physics.  
   Therefore, our  results are  equally-applicable to any other localized  objects on asymptotically
   Poincare-invariant  space, such as classicalons.    
  By adding a requirement  that in large mass limit the quantization must approximately   
 account for classical results, we derive an universal quantization rule applicable to 
 all classicalons (including black holes) in arbitrary number of dimensions.     
   In particular, this implies,  that black holes cannot emit/absorb arbitrarily soft quanta.   The effect has phenomenological  model-independent implications for black holes and other classicalons that may be created at LHC. 
 We predict,  that contrary to naive intuition,  the black holes and/or classicalons, will be produced in form of fully-fledged quantum resonances of discrete masses, with the level-spacing controlled 
 by the inverse square-root  of cross-section.  
 
%  The classicality sets-in  in form of closer and closer level-spacing 
 % for the increasing mass, but continuous masses are never achieved.   
 
\noindent

\vskip .1in
\noindent

\section{Introduction and Summary}

  Theoretical  realization of possibility of low quantum gravity scale \cite{add, aadd}, 
  opened a prospect of potential  experimental studies of quantum gravity at  the particle 
  accelerators.    One of  the model-independent signatures of strong gravity at LHC  was  predicted\cite{aadd} to be formation of micro black holes.  For  subsequent studies of 
  black hole formation in colliders see\cite{BH} .    
 Of course,  experimental searches are unimaginable without at least qualitative theoretical understanding of properties of micro black holes, to which we often mistakenly extrapolate  
 idealized classical properties.  In our previous  paper \cite{gia-cesar-slava} we have 
 shown  that micro black holes must be quantized,  and have provided evidence 
 that the quantization rule in high mass limit must reproduce the area quantization rule conjectured 
 in pioneering work of  \cite{bekenstein,slava,BM}. 
  In the present paper we shall give an exact prove of  black hole mass quantization from the first principles of quantum physics, and derive 
  universal quantization rule, that  for high levels reproduces the particular  result of \cite{bekenstein, slava, BM}.

 The purpose of the present  work is to unambiguously show that 
 black hole masses (as masses of all the other bound-states  in the quantum world) are quantized, and this fact determines  the properties  of micro black holes that may be accessed at LHC.

  As it is well known,  classically, the black holes are strongly-gravitating objects that can be produced  as a result  of a gravitational collapse of an arbitrary form of energy that can be localized within its own Schwarzschild radius.   In quantum field theory framework, any form of energy can be described  as  some superposition of elementary particles that correspond to quantum degrees of freedom existing in the theory. When the energy is so large that the 
corresponding classical Schwarzschild radius  exceeds all the relevant quantum length-scales in the problem, 
the configuration becomes effectively {\it classical}. 
 In the particle language, classicality can be understood as
the occupation number of particles involved in the configuration being  large.  The constituent particles 
that make up a given classical object, can be either on-shell or off-shell,  but their number  is always large, and this is what defines classicality. 
  For example,  for  time-dependent configurations, such as gravity or electromagnetic waves, the constituent particles (gravitons or photons)  are mostly on-shell.    Contrary, in the static  
configurations, such as  a black hole or an extended strong magnetic field, the constituents are 
off-shell.  We can say,  that such configurations are made out of longitudinal gravitons or photons of large occupation number. 
  In both cases, classicality can be understood as the number  of constituent quantum particles, 
  being large, $N  \, \gg \, 1$.   But, since underlying laws of physics are quantum, the configuration is always also quantum.   For any finite $N$ and finite size, the energy of any such configuration must be quantized.  
 This is true about  a hydrogen atom, a neutron star  and a black hole. 
 
   Despite of the above obvious point, black holes  are still often perceived as something exceptional from this rule.   The reason for this misconception is perhaps our intuition based on the fact that in exact-classical limit 
 $(\hbar \rightarrow 0)$, black holes are described by solutions that  are fully characterized by 
 very few parameters, such as the mass, charge and an angular momentum. 
 This fact  is sometimes referred to as a black hole no-hair theorem \cite{nohair}, and it is a powerful tool 
 for understanding certain approximate properties of real black holes. Nevertheless,  it 
 represents  an idealized approximation,  that can never be realized in real systems existing for the finite time.   
 This idealized property, valid in an exact classical limit,  we  mistakenly often extend to real black holes. 
  
 %   One possible reason why we may  think that black hole mass can assume a continuum of values is, that according to no-hair property, the black hole mass is independent of the precise  form of energy that formed it.    Since the energy of an initial state  is not quantized, so must be 
%the mass of a black hole.  This reasoning, however, is  mistaken, since although energy in general is not quantized,  the energy of any localized bound-state is.   

 The above misconception has most severe consequences for the experimental searches of micro black holes  at LHC, because it suggest that in such searches we should look  for semi-classical objects 
  that evaporate {\it democratically}  in all the light particle species with nearly a thermal spectrum. 
  In reality, the bottom of a black hole tower that may be reached by LHC, will look nothing like this. 
  The first resonances that we may be lucky to observe will be fully legitimate quantum resonances, with quantized mass spectrum and no obvious reason to  decay democratically in all the species.   
  
 Quantization of their masses 
  we shall prove unambiguously below.    The issue of democracy was discussed in detail in \cite{democ}. As it was shown there, by unitarity and CPT the democracy  must be a mass dependent  property,  with  the lightest members of black hole tower   exhibiting  the lowest level of it.  Further refinements of these results  are beyond the scope of the present work. 
  
 %  Quantization of the black hole masses  was conjectured earlier in \cite{Bekenstein, slava, BH}
  % in the form of the requirement that the black hole area must include integer number of Planck area units, $L_P^2$.  
       
 In \cite{gia-cesar-slava}, we have argued that quantization of black hole masses follows from the 
 fact that Planck length $L_P$ is the shortest resolvable length scale in gravity, and is necessary 
 for consistency of  the transition between elementary particles and black holes  required by self-completeness of gravity\cite{gia-cesar}. 
 We showed,   that assuming saturation of the holographic bound on information storage 
 exactly reproduces the area quantization rule conjectured in \cite{slava, bekenstein}.  
 We have further studied  implications of mass quantization for  LHC searches of black holes. 
   
    In the present paper,  we shall prove inevitability of black hole mass quantization from the first principles of quantum theory  and Poincare-invariance, with no other assumptions.  
   Because of this generality, our proof is not specific to black holes and holds 
   for arbitrary localized states.   
   
   Other phenomenologically-interesting objects that automatically fall within 
   our proof  are recently-suggested {\it classicalons} \cite{class}. 
   They represent generalizations of black holes 
   for non-gravitational theories that include  bosonic fields  (self)sourced by energy.   
 Interest in these objects is particularly motivated by the idea  that they can be an intrinsic part of the standard model  unitarizing the scattering of longitudinal $W$-bosons via classicalization.   In such a case, of course, proof of quantization is  very important for the proper  understanding their collider signatures.  Our results show,  that $W$-classicalons, just like micro black holes,  must manifest themselves  in form of a  tower of quantum resonances  of  discrete masses.

       Due to generality of our argument,  under a  {\it black hole}  we shall mean an arbitrary configuration characterized by a mass $m$ and a localization  radius $r_*(m)$.    At no point in our proof we shall use the
 geometric characteristics of a black hole.  For example, we won't rely on any particular relation 
 between  $r_*$ and $m$.  For us, it will be sufficient  that  a characteristic size of a black hole 
 can be defined for the cases of interest and is a finite quantity for finite $m$.  
  Our reasoning will be performed from the point of view of an asymptotic observer 
  for which a black hole is just another state of a finite  mass, and for which its precise geometric structure is completely  unimportant. 
 
 Of course, since we work in quantum field theory framework, at the end of the day, all the states are quantum (pure or mixed  doesn't matter)  
and must be characterized by a corresponding wave-function $\psi$ describing a state in the 
Hilbert space. 
 The statement that a given state has a finite mass and a localization radius, 
 implies that the corresponding wave-vector  $|\psi\rangle$ has a finite norm. 
   Thus, we are dealing with a localized wave-packet rather than a plane-wave.  
 We wish to show, that the mass assumed by any such state, must be quantized. 

 We shall first consider a situation with a stable state, and then take into the account the effect of 
 decay. 
 
  We shall consider  two field theoretic arguments showing that allowing a continuum of masses 
 leads us to a contradiction, and thus, the  black hole mass must assume discrete values.  
 
   We shall summarize our argument is one line.  The key point is Poincare invariance of the 
   asymptotic background.  On such a background, any  $|in \rangle$  and $|out \rangle$ states  in a scattering process 
   can be labeled  by quantum numbers describing irreducible representations 
   of the  Poincare group.     
    Consider a scattering process  that  takes an  in-state  $| in \rangle$ into a final states 
    $|f_m \rangle$,  that is characterized by a Poincare-invariant parameter $m$. 
    Let the rate of this transition be  $\Gamma_{in\rightarrow f_m}$.  The total rate of the 
    process  is then obtained by summing up over $m$, 
      \begin{equation}
  \Gamma_{total}\, = \,   \, \sum_m  \Gamma_{in\rightarrow f_m} \, .
   \label{rate}
  \end{equation}  
 All the phase space integrations over the Poincare-non-invariant parameters (such as momenta of final particles) are already included in  $\Gamma_{in\rightarrow f_m}$.
  So the remaining sum can only take place over Poincare-invariant parameters, such as 
  masses, spins,  and internal quantum numbers.   Allowing these parameters to assume continuous values on any interval will inevitably result into an infinite transition rate, and would make no sense. 
  
   It is important to appreciate the power of the above simple argument, which only relies 
 on Poincare invariance of an asymptotic background, and is therefore completely insensitive    
 to the short-distance (or high energy) properties of the interaction that is responsible for 
 the transition.  This is why, the masses of localized objects must be quantized regardless 
  what  one is willing to  imagine about  short distance properties of quantum gravity. 
  
   From the first principles of quantum physics, we  thus prove that masses of black holes or any 
   other classicalons must be quantized. 
    However, to uncover the precise quantization rule we need more input.  
   We use as such an input the requirement that in $m \rightarrow \, \infty$  limit the quantization rule should reproduce the classical relation between   $r_*$  and  $m$.   We then obtain  the following {\it universal}  quantization rule  that is accurate in this limit, 
   \begin{equation}
      (m r_*)   =   N  \, .
      \label{rule1}
      \end{equation} 
Notice,  that  for a four-dimensional  Einsteinian  black hole this  exactly reproduces earlier conjectures \cite{slava, bekenstein},  of quantization of black hole area in $L_P^2$-units.
However, the rule given by (\ref{rule1})  is more generic  and universally applies to a quantization of arbitrary  classicalons  (including black holes) in arbitrary number of dimensions. 
 In the light of ref. \cite{gia-cesar-alex}, the  above rule has a  clear physical interpretation. 
 It was shown there that  certain properties of classicalons  (and in particular black holes)  can be 
 understood by realizing that they represent configurations  of large occupation number  
 of the soft  bosons of  wavelength $r_*$, with the number given by (\ref{rule1}). 
  The quantization rule, then simply can be understood as trivial consequence of the quantization 
  of the occupation number.   For large $N$ contribution coming from shorter wave-length
  particles is less and less important, and the rule becomes more and more accurate. 
      This result is also  understandable  in the light of an underlying holographic connection between different  classicalizing theories  explained  in \cite{gia-cesar-alex}, and provides extra evidence for  this connection.

      Finally, we shall show that the mass-quantization  of sharp black hole resonances  is insensitive to the possible fine level splitting of underlying micro-states, and is given by 
      the scale of the production cross section
   \begin{equation}
   \Delta m \, = \, 1/\sqrt{\sigma} \, ,
   \label{crossrule}
   \end{equation}
   even if the underlying "true" mass levels are extremely densely spaced.  
   In phenomenological terms the above rule tells us the following. Let us imagine that in a scattering process 
   we can resolve a  sufficiently narrow black hole resonance of mass $\bar{m}$  that is  produced with a  cross section  $\sigma$. The rule then tells us that a neighboring  resolvable  resonance  
will be shifted approximately  by  (\ref{crossrule}).   This rule does not preclude the existence of the 
substructure of levels,  but at each resolution level the same rule should apply to the level spacing.

    We shall now turn to more detailed discussion. Before doing so,  let  us briefly comment on notations.  Throughout the paper  we shall denote by $\bar{m}$ the masses of resonances, 
    whereas the masses of exact mass eigenstates will appear without bar,  $m$. 
     Also in some obvious places we shall drop coefficients of order  one.

\section{The Need for  Mass-Quantization} 
  
 \subsection{Infinite Creation Rate for Black Holes with Non-Quantized Masses} 
 
  First argument shows that creation rate of black holes with continuum masses would be infinite, 
  and thus such theories cannot make physical sense. 
 
  To see this, let us assume,  that the black hole spectrum in some interval of masses is continuous.  
Let us pick one member from this continuum of mass $m$.  As said above, because a black hole 
is a localized state,  and can be produced with finite probability in a collision process, 
it must have  a finite norm in Hilbert space.   We shall show that relaxing the condition of the finite norm will not change our conclusions, since only the finite-norm states can be created in scattering experiments. 

Consider thus any scattering process that  creates such 
an objects in a final state.  For example,  this can be a process in which some initial quantum particles scatter into a black hole plus some other particles, 
\begin{equation}
 {\rm initial~particles} \, \rightarrow \,  {\rm  BH_m}  \, + \, {\rm some~final~particles} \, , 
 \label{process1}    
 \end{equation}
 or  simply a process of a black hole pair-creation out of some initial state,  
 \begin{equation}
 {\rm initial~particles} \, \rightarrow \,  {\rm  BH_m}  \, + \, {\rm \overline{BH}_m} \, .  
 \label{process2}    
 \end{equation}
 Here  ${\rm  BH_m}$  and  ${\rm \overline{BH}_m}$ stand for a BH of a definite mass $m$ and its charge-conjugated state respectively. 
 Basically, any process in which a black hole  can appear in a final state 
would do the job.  
 
 In any sensible field theory an amplitude of such a process, 
\begin{equation} 
A  \, \equiv \,  \langle initial~particles | BH_m \, + \,  something \rangle \, ,
\label{amplitude}
\end{equation}
 must exist   and be non-zero  at  least for  some initial states. 
 
  Obviously, for many  initial states, such a transition may be extremely suppressed, but 
  this is no-problem for us. As long as the amplitude is finite, we can build our argument. 
  The finiteness of such an amplitude follows from the fact that ${\rm BH_m}$ corresponds 
 to a finite norm vector in Hilbert state.  If norm were infinite, then it would mean that 
 probability of creating ${\rm BH_m}$ is zero, and only a distribution (a resonance composed 
 out of continuum of mass  eigenstates)  can be created.   
  This is obvious from the fact that the  expectation value of any Hermitian operator ${\mathcal M}$ over such a state would vanish 
 \begin{equation}
    \langle  BH_m | \, {\mathcal M} \,  | BH_m \rangle \, = \, 0 \, .
    \label{massmeasure}
    \end{equation} 
%  Moreover,  if amplitudes (\ref{amplitude}) are always zero, then effectively in such a field theory black holes 
 % do not exist, since they cannot be prepared within a finite time from any  reasonable initial state.  
 %In particular, such black holes cannot be produced in proton-proton collisions at LHC.  
% We are not interested in such theories.  
 
  Let for some  fixed energy of an initial state,  $\sqrt{s}$,  the total rate of such a process in which the black hole of a particular mass appears,  be $\Gamma (m)$.  By analiticity of  the scattering  amplitudes, the function $\Gamma(m)$ must be 
  non-singular at least in some finite interval of masses
  \begin{equation}
  m_0 \, < \, m \,  < \, m_0 + \Delta m \, .
  \label{interval}
  \end{equation}
   This must be true regardless of  particularities of the dynamics that is responsible for 
   black hole creation.   
   
 Since this is the most crucial step, let us elaborate on it. 
 Continuity of the production rate  in the processes  (\ref{process1}) means, that if 
in the collision at center of mass energy $\sqrt{s} \, > \, m$, one can produce a black hole 
of mass $m$  with four momentum $p_{\mu}$ and a particle of some four-momentum 
$q_{\mu}$, the amplitude  of a similar process with  infinitesimally close value of mass
($m' = m+\epsilon$) and some respective values of four-momenta ($p_{\mu}' , q_{\mu}' $) must be infinitesimally close. 

  After integrating over final four-momenta the resulting rate 
$\Gamma(m)$ must be a continuous function of $m$, meaning that 
$\Gamma(m+\epsilon) \, \rightarrow  \, \Gamma(m)$ for $\epsilon \, \rightarrow \, 0$.  
The same is true about the process of black hole 
pair creation  (\ref{process2}). 

 Because of this continuity, one cannot assume, for instance, that a black hole of mass 
 $m$ can only be  created on a mass-threshold.  This would inevitably violate unitarity and 
 the asymptotic Poincare-invariance in such interactions.  
 
  In other words, once we make the assumption that  black holes can be created in particle collision 
  process in a theory in which particle interactions respect unitarity,  we are forced to assume 
  that black hole creation processes also respect the same consistency rules. 
  In a consistent particle theory we cannot think of black holes as special objects  that are not subject to common rules of the game, since any violation of unitarity rules by black hole processes 
  would contaminate  also the interactions of ordinary particles. 
  
   The only way to give black holes a special status, is to completely isolate them, and eliminate  possibility of their  production in ordinary particle processes. Such complete elimination is hard to 
   understand in the view of $CTP$-invariance of the theory, according to which a pair-production of any 
  finite energy/norm state at high enough energy must be allowed by all possible super-selection rules.   For example,  since a black hole and its charge conjugated state can annihilate 
  into gravitons, the inverse transition must also be allowed.  
    But even if one somehow can imagine a consistent loophole out of this argument, 
   we are not interested in theories in which black holes cannot be produced in particle collisions.  
   Speaking the least, such black holes will never be observed in collider experiments. 
   
    Let us thus proceed by accepting the fact that continuity of $m$ implies continuity of $\Gamma(m)$ at least on some interval (\ref{interval}).   
  Let the minimal value of $\Gamma$ on this interval be $\Gamma_{min}$.  
  In order to obtain the total rate of emission within this interval we have to sum up over all the members of the continuum, which obviously will give an infinite result, 
  \begin{equation}
  \Gamma_{total}\, = \,   \, \sum_m  \Gamma(m) \, = \, \lim_{n\rightarrow \infty} \, \sum_{k=0}^{n}  
  \Gamma(m_0 +  k \Delta m/ n) \,  > \, \lim_{n\rightarrow \infty} \, n  
  \Gamma_{min} \, = \infty   \, . 
  \label{total}
  \end{equation}  
  Where,  in order to make the divergence of the sum explicit, we have discretized the interval
 of masses by $n$ steps and then have taken $n$ to infinity.  
   
    Notice, that the above sum is not a simple integration over $\Gamma(m) dm$ with an infinitesimal measure, which of course would give a finite result.  The reason is,  that we are summing up over an infinite number (continuum) of final states, each with {\it finite} norm.
           The above result summarizes our point. The equations (\ref{amplitude}),  (\ref{massmeasure})  are incompatible with masses being continuous  on any interval (\ref{interval}), since such continuum leads  to an infinite production rate, which makes no sense. 
 
    In a consistent  $CPT$-invariant quantum field theory anything that has finite energy and can fit within a finite size box, can be pair produced with a non-zero probability, provided  one pumps enough energy within that box.

           Relaxing any of the two conditions  (\ref{amplitude}) or  (\ref{massmeasure})  would mean  eliminating the black hole states with finite  mass and a finite norm from the  Hilbert space of the theory.

          As said above,  we are not interested in such a situation.   
   The only less radical step is to assume, that the black holes of definite mass do exist, 
   but the possible values of the mass are discrete.   This makes perfect sense, and is fully compatible  with everything  we know about quantum physics.         
  
  \subsection{Infinitely Strong Couplings Mediated by Virtual Black Hole States with Continuous Masses}

   The second argument  is based on processes mediated by virtual black hole states. 
  Again,  we can show that if the black hole masses are not quantized, integrating out such virtual states would lead to effective interactions of infinite strength among the low-energy fields. 
  
 Indeed, consider a black hole state of a definite mass $|BH_m\rangle$.   If such a  black hole state with a well-defined mass and a finite norm exists,   then in any sensible quantum field theory it should also appear in form of  an intermediate state in some transitional process. That is, an amplitude,  
 \begin{equation}
 \langle initial~particles | BH_m\rangle \langle BH_m | final~particles \rangle  \, \neq \, 0
 \label{transit}
 \end{equation} 
 must be non-zero at least for some initial and final states.  
  Then,  by integrating out such virtual black hole states one should generate some effective operators  describing interactions among the light fields at low energies.  
   Let an operator generated by integrating out the above  member  of  the black hole continuum of a given mass $m$ be,    
  \begin{equation}
   \Phi_1 ... \Phi_{K} \, ~  A(m) \,, 
 \label{operator}
 \end{equation} 
 where $\Phi_j\, , ~ j =1...K$ stands for some light field of different  possible spins and other quantum numbers.  The coefficient  $A(m)$ controls the strength of the interaction.
  Now again, in any sensible field theory with finite norm states of mass $m$, at least some 
  low energy operators must be generated by integrating these states out. This means that 
  $A(m)$ must be finite and non-singular at least in some interval  (\ref{interval}) . 
   Moreover, we can always choose the interval in such a way that  $A(m)$ has a definite sign on it, 
   and a lowest value $A_{min}$. 
  Then, by integrating over all possible virtual black holes on such an interval we would generate 
  a low energy  interaction of an infinite strength, 
    \begin{equation}
   \Phi_1 ... \Phi_K  \sum_m  \, A(m) \, = \, 
  \Phi_1 ... \Phi_K   \lim_{n\rightarrow \infty} \, \sum_{k=0}^{n}  
  A (m_0 +  k \Delta m/ n) \,  = \,  (\Phi_1 ... \Phi_K) \,  \infty \, .  
 \label{operatorinfinite}
 \end{equation} 
   Thus exchange of continuum of virtual black holes, no matter how suppressed for individual ones, 
  would generate low energy interactions of an infinite strength. 
  
   Of course, one can assume that 
  summing over all possible black holes may result in a  destructive interference, but if we had to rely on 
 such interference,   this would be the end of low energy physics as we know it, since in order to compute any low energy process, one would need to rely on cancellations among the virtual superheavy objects that separately give infinite contributions.  
 Of course, on the basis of first principles we cannot exclude such a miraculous cancellation of contributions 
in the low energy effective operators. But, even if this happens, this cannot save the blow-up of a 
scattering amplitude demonstrated in the previous section, since the black holes there appear in the final states, and all the contributions to the production rate are positive.  

  We thus conclude,  that existence of stable black holes (or any other localized bound-states) 
 with continuous  values of masses is impossible in a consistent quantum framework that allows 
 quantum production of these objects in the final states.   An immediate consequence
 of this fact is that the masses of extremal black holes, e.g., RN ones,  must be quantized. 
 Of course, such quantization follows from quantization of an electric charge. 
 But,  we are arriving to the necessity of quantization from completely independent  side, which makes mass quantization  a necessity even if  charge could take a continuous value. 
 % (e. g. as it happens for a magnetic monopole in the presence of a $\theta$-term). 

  \subsection{Why the Infinite Norm Cannot Undo Quantization of Masses}
 
  Can a  black hole state in the Hilbert space have an infinite norm? 
  As we shall now show,  this will not change anything in our conclusion, since even 
 if one formally prescribes to the black hole states in the Hilbert space  
  an infinite norm,  the states that can be physically produced and observed will anyway have a quantized mass spectrum. 
 
  To see this, let us assume that black hole states of definite mass (call them  $|BH_m\rangle$)
have no finite norm, and thus are allowed to have a continuum of masses $m$. 
  To form a physically-accessible Hilbert sub-space, they should be at least $\delta$-function- 
  normalizable, 
 \begin{equation}
  \langle BH_{m'} ||BH_m\rangle \, = \, \delta(m'-m) \, ,  
  \label{dnorm}
  \end{equation}
 and (if needed,  together with some additional states)  can be used to form an orthonormal complete set.  
  
  Obviously,  in any local physical process, we can only produce a finite-norm superposition of the states 
  $|BH_m\rangle$. Let us denote these observable finite-norm states by 
   $|\bar{BH}_{\bar{m}}\rangle$.  Of course, by default, these finite-norm observable states are {\it distributions}, and cannot have a definite mass. Each of them represents a superposition (wave-packet) of 
   the definite mass (but infinite norm) states, 
  \begin{equation}
 |\bar{BH}_{\bar{m}}\rangle \,   =  \,  \int \,  dm  \, \psi_{\bar{m}}(m) \,  | BH_m\rangle \, ,  
  \label{superposition}
  \end{equation} 
 where  $\psi_{\bar{m}}(m)$ define the relative weights by which an $m$-th black hole state 
 enters in a given  $\bar{m}$-th superposition. 
 Because of orthogonality properties (\ref{dnorm}), 
 we have 
   \begin{equation}
   \langle \bar{BH}_{\bar{m}} ||\bar{BH}_{\bar{m}}\rangle \,   =  \,  \int dm \,  \rho_{\bar{m}}(m) \, , 
  \label{distr}
  \end{equation} 
where $\rho_{\bar{m}}(m) \, \equiv \, |\psi_{\bar{m}}(m)|^2$ is a spectral density function. 
  Of course, all the states $\bar{BH}_{\bar{m}} $ are unstable, but a well-defined resonance 
  exists only if the spectral density function is peaked about some value $\bar{m}$, with the 
  width satisfying 
  \begin{equation}
   \Delta(\bar{m}) \, \ll \, \bar{m} \, .
   \label{resonancecond}
   \end{equation}
  If these conditions are not met, 
  the state is so broad, that not even approximately shares any properties with black holes. 
  Obviously, this is not the case of our interest.  
  
   We thus focus on the case when   (\ref{resonancecond}) is satisfied. 
   
   Let us now ask, if such resonant states can ever come in form of a  continuum?  
     To see that this is not possible, we can repeat the infinite production rate argument but now applied to   the  $\bar{BH}_{\bar{m}}$-states.  These states are labeled  by 
   $\bar{m}$.    Let us ask, if parameter  $\bar{m}$ can take a continuum of values. 

    Since the state $BH_{\bar{m}}$ is normalizable, the production rate $\bar{\Gamma}(\bar{m})$ must be finite and non-singular on some interval,
     \begin{equation}
  \bar{m}_0 \, < \,\bar{m} \,  < \, \bar{m}_0 + \Delta \bar{m} \, .
  \label{intervalbar}
  \end{equation}
 Let the minimal value of $\bar{\Gamma}(\bar{m})$ on this interval be $\bar{\Gamma}_{min}$.  
  Then again summing  up over all the members of the continuum,  will give an infinite result, 
  \begin{equation}
  \bar{\Gamma}_{total}\, = \,   \, \sum_{\bar{m}} \bar{\Gamma}(\bar{m}) \, = \, \lim_{n\rightarrow \infty} \, \sum_{k=0}^{n}  
 \bar {\Gamma}(\bar{m}_0 +  k \Delta \bar{m}/ n) \,  > \, \lim_{n\rightarrow \infty} \, n  
 \bar{\Gamma}_{min} \, = \infty   \, . 
  \label{totalbar}
  \end{equation}  
This result illustrates that the number of distinguished finite norm states,  both the mass eigenvalues as well as sharp resonances, must be quantized. 
     
     This fact may come as a surprise, because naively one could argue in the following way. 
  Let us start with a theory in which a number of  finite-norm mass-eigenstates (or sharp resonances) is discrete. 
   Now  let us pick up two such distinct mass eigenstates denoted by   $|m \rangle$ and $|m' \rangle$. 
   Then one may expect that  in a scattering process starting  from some initial state, $\langle i |$, one should be able 
   to produce  a normalized superposition of the two states,
   \begin{equation}
  |\alpha \rangle  \, \equiv \,   \cos (\alpha) |m \rangle \,  +  \, \sin(\alpha) |m' \rangle \, , 
   \label{alphastate}
   \end{equation}  
    plus some additional state,  which we shall denote by  $|f\rangle$.  
      Of course,  the state 
    $|\alpha \rangle$ will evolve in time, because the two mass eigenstates will oscillate. 
    But, one could argue, that the  mass difference can be taken arbitrarily small,
    so that the oscillation period $\tau \sim (m-m')^{-1}$ can be made arbitrarily longer than 
    $m^{-1}$.   
    In this way, the state $|\alpha \rangle$ can be made into an arbitrarily-sharp resonance, 
    with the lifetime exceeding all the other time scales in the problem. 
     In such a case, the final state can be treated as direct product, 
   \begin{equation}
  |\alpha, f \rangle  \, \equiv \,  |\alpha \rangle \,  \otimes  \,  |f \rangle  \, . 
     \label{truefinal}
   \end{equation}  
     Can the parameter $\alpha$ take continuum values in such  process? Our argument says that this is impossible.  Indeed, if the transition amplitude $\langle in| \alpha, f\rangle$ where finite, 
and continuous  on some interval  
\begin{equation}
\alpha_0  \,  < \,\alpha  \, <   \alpha_0 + \Delta \alpha_0 \, , 
  \label{alphaint}
  \end{equation}
 The total rate that would be obtained by summation  over this interval would be infinite. 
 
  But,  what is the fundamental reason for  the discreteness of $\alpha$? 
  The reason, in fact, is Poincare-invariance.   By  Poincare-invariance,  the parameter 
  $\alpha$ can only depend on {\it Poincare-invariant} characteristics of the state 
  $|f\rangle$.   These are,  eigenvalues of Poincare Casimir operators (masses and spin) 
  and/or  internal quantum numbers that are Poincare-singlets.  Since number of particles in $|f\rangle$ is finite 
  and discrete,  all these characteristics are discrete. Thus, the  summation over 
  the label $\alpha$ is always discrete.

    Impossibility of creating continuous superpositions of two states can be easily visualized 
    on the following simple example. Consider a theory in which an initial state, 
    let us say a heavy fermion $\chi$, decayes into a superposition of two nearly degenerate 
    massive fermions, $\nu_{\alpha} \, \equiv \nu_m  \cos(\alpha) \, + \, \sin (\alpha) \, \nu_{m'}$ and a light scalar  $\phi$, 
     \begin{equation}
      \chi \, \rightarrow  \, \nu_{\alpha} \,  + \, \phi  \, .
    \label{decay}
    \end{equation}
    Can one write down an effective vertex, that would allow in this process to produce 
    continuum of different $\alpha$-superpositions?  The answer is no, because of Poincare-invariance.   The corresponding vertex has a form 
      \begin{equation}
      \chi \, \nu_{\alpha}  \phi 
    \label{vertex}
    \end{equation}
  and the only way to make different values of $\alpha$  possible in this process, is to make $\alpha$ sensitive to the Poincare-invariant continuous parameters that characterize final state of $\phi$. But 
  because of Poincare symmetry, such continuous parameters do not exist.   In the absence 
  of Poincare  invariance, we could make $\alpha$ being explicitly dependent on 
   four-momentum of $\phi$, but Poincare-invariance forbids such dependence indicating that 
   $\alpha$ can only depend on eigenvalues of Poincare Casimir operators, such as four-momentum square $p_{\mu}p^{\mu}$, or other operators commuting with the Poincare group. 
   For any given $\phi$, the set of such parameters is fixed and is discrete.  So there is no way to 
   vary  this values as functions of  four-momenta of the final states.

 As a result,  different $|\alpha\rangle$-states can only be produced if we also make 
 $\phi$-state $\alpha$-dependent.  We can achieve this for example either by introducing 
 different $\phi_{\alpha}$-fields and writing set of interactions
   \begin{equation}
      \chi \, \sum_{\alpha}  \nu_{\alpha}  \phi_{\alpha}  \, , 
    \label{vertex2}
    \end{equation}
or introducing operators that couple to different powers of $\phi$, say,  
   \begin{equation}
      \chi \, \sum_{\alpha}  \nu_{\alpha}  \phi^{\alpha}  \, . 
    \label{vertex3}
    \end{equation}
Obviously in both cases, the set of possible operators is discrete.     

    So the observable black hole states, must form a discretuum.  
 
   The above is a manifestation of a simple, but very deep, quantum-mechanical truth: 
 {\it Resonances are normalizable states built out of continuum,  because of this property they come 
 in discrete numbers!}. 
 
  In other words, the continuum of resonances makes no quantum-mechanical  sense.

% \subsection{Taking into the account thermal-type instability}  

 The above consideration can be straightforwardly applied  for accounting other types of instabilities, such as  the Hawking  thermal instability for a black hole.  
  Instability can be easily accounted by treating a black hole state as a distribution 
  (a resonance), of spectral density $\rho_{\bar{m}}(m)$.   Then we can automatically repeat the 
  previous arguments, and conclude that $\bar{m}$  must be  quantized.

 \section{How Quantized?}
 
  We have concluded  that masses of localized  objects, such as  black holes and other classicalons,  must be quantized.   
  Can we derive a quantization rule? 
  To do this we need a more detailed information about the objects in question. 
  
 In order  to derive  an approximate quantization rule, we can proceed in the historic spirit of 
  quantization, by demanding that the quantization rule for large $m$ (meaning $m \gg L_*^{-1}$) 
  should approximately reproduce the results of semi-classical computation obtained in the approximation of a  continuum mass.   
  
   Interestingly,  for classicalons (and in particular black holes) it turns out to be easier to guess an universal rule and then discuss  its supporting evidence in concrete cases.  
  In making such a guess,  we shall rely on the results of ref\cite{gia-cesar-alex} , which showed  that  (seemingly mysterious) physical properties of  classicalons can be well-understood  if we realize 
  that  these objects represent superpositions of  soft (wavelength $\sim r_*$) quanta 
  with  occupation number given by (\ref{rule1}). 
  
   The quantization rule then follows from the following simple consideration.  For large $m$ 
   ($N \, \gg \, 1$) the main contribution to the mass (and other properties) of the configuration 
   comes predominantly from the wavelengths $\sim r_*$.   Contribution from short wave-lengths 
   is negligible.  So the quantization rule then can be understood as the quantization of the occupation number of these cold bosons.  
  This rule automatically accounts for  all the existing classical limits. 

 In each concrete case,  we can gether more supporting evidence for it, which we shall now do, 
 separatelly for the black holes and other classicalons. 
        
    \subsection{Black Hole Mass-Quantization} 
  
     One of the well-known results obtained in the semi-classical approximation of continuous mass is , that black holes decay via emitting a thermal Hawking radiation of temperature $T \, = \, r_*^{-1}$.  
   Thus,  in this approximation  a black hole of mass $m$ emits predominantly the  quanta of frequency 
   $T$.     We therefore require that the quantization rule should accommodate this result, meaning that for large mass (small $T$), the separation between the levels must
   decrease as $T$. 
   
  In order to accommodate the semi-classical limit  for large masses,  the quantization rule should   
 allow a  black hole of mass  $m$  to go to a lower level $m'$ by emitting a quantum of some massless or massive (subject to Boltzmann-suppression)  particle, $\gamma$ of energy $T$,   
  \begin{equation}
 {\rm BH}_m \, \rightarrow   {\rm BH}_{m'}  \, + \,  \gamma  \, .
  \label{emission}
  \end{equation} 
  For large $m$ (small $T$) the back reaction is higher order  in  $T/m$, 
 and can be  ignored and the level difference is approximately given by 
 $m \, - \, m'  \, = \,  T$.   Now, even without  knowing a concrete dependence between 
 $m$ and $T$, but solely relying on the fact that in each elementary emission the back reaction 
 is small, we can conclude that the number of quanta a black hole needs to emit before substantially reducing its mass is 
 \begin{equation}
   N \, = \, m/T \, = \, mr_*  \, . 
 \label{ruleT}
 \end{equation}
 So whatever the quantization rule is, for large $N$ it has to reproduce this information.  

     In order to translate this in terms of mass quantization, we need a concrete relation 
   between $r_* (T)$ and $m$, which is  more model dependent, e.g., depends on  number of extra dimensions.  But, the rule (\ref{ruleT}) is unchanged. 
   
  For example, for  Einstenian gravity in four dimensions,  we have   $r_* \, = \,  L_P^2/m$.
  If we make a further reasonable assumption that the transition probability is maximal between the 
   neighboring levels,  the resulting quantization rule,  in the leading order in $N$,  comes out to be 
   \begin{equation}
   m = \sqrt{N} /L_P \, , 
   \label{quantrule}
   \end{equation}  
   which exactly reproduces conjecture of \cite{bekenstein,slava, BM}.  Also note, that when applied  to black holes in $4+d$ dimensions, the rule (\ref{rule1})  reproduces the quantization rule  obtained in \cite{gia-cesar-slava} of $2+d$-dimensional black hole area
  $A \, = \, r_*^{2+d}$   
    in units of 
 $1+d$-dimensional fundamental Planck area,  $L_{4+d}^{d+2}$.  Or in terms of masses this quantization rule reads,    
   \begin{equation}
   m \, = \,  N^{{1 \over 2+d}} \,  /L_{4+d} \, . 
   \label{quantruled}
   \end{equation}

    An alternative estimate of the quantization rule, which gives the same result, 
comes from the requirement  that a black hole creation cross-section in two-particle collisions at very high center of mass energies $m$, should approach a geometric cross section set  
   by the area of a Schwarzschild black hole of mass $m$ \cite{gt, bh2, bh3, bh4, bh5, bh6}, 
   \begin{equation}
    \sigma_{m} \, = \,  r_*(m)^2 \, .  
  \label{cross-section}
  \end{equation}
 This process, is accompanied by creation of additional soft quanta, so that  schematically we can write down, 
 \begin{equation}
 \gamma + \gamma \, \rightarrow  \, BH_{m +\Delta m} \, + \, \gamma  \, ,
 \label{softemition}
 \end{equation}
 where $\gamma$ stands for some initial and final particles. 
 Since the scattering takes place at a macroscopic distance $r_*$,  the characteristic momentum transfer is $\sim 1/r_*$,  which limits the typical momentum of the final particle from above. 
 Obviously, if we want the total cross section of such a process,
  \begin{equation}
   \sigma_{total} \,  = \sum_{\delta m} \,   r_*^2 (m+\Delta m) \, ,   
  \label{cross-section}
  \end{equation}
  not to exceed (at least significantly)  
 the geometric cross-section, we cannot allow the values of $\Delta m$  much smaller than 
 $1/r_*(m)$.   Thus, again we have to admit that the level separation is $\Delta m \sim 1/r_*$, which again 
 leads us to the quantization rule that approximately is given by relation  (\ref{rule1}). 
 
  \subsection{Quantization of Classicalon Masses}

    Another category of localized finite mass configurations, that generalize notion of 
    black holes, to a more general class of theories, are classicalons \cite{class, gia-cesar-alex}. 
    In fact, as shown, black holes are simply a particular form  of classicalons. 
     The crucial unifying property of classicalons is  that their size $r_*$ grows with their mass. 
    The mass (and thus size) appear at the level of the classical equations of motions as an integration constant that like a black hole mass, can assume a continuum of values. 
     But this continuity, just like in a black hole case is an artifact of classical approximation. 
  Just like black holes, classicalons can be viewed as bound-states of many bosons, with
  characteristic wave-length $r_*$.   For total energy $m$, the occupation number is   
  obviously  given by (\ref{rule1}).
 % \begin{equation}
 % N \sim r_*m \, .
  %\label{nquanta}
  %\end{equation}
   For a concrete case of black holes the role of $r_*$ is  played 
  by a gravitational Schwarzschild radius, but  relation  (\ref{rule1})  holds also for other classicalons. 
  
    From all these features, it is obvious that the classicalon masses must be quantized according to the same rule.  
    To formulate a  quantization rule only in terms of mass and $N$,  we need to know dependence  of  $r_*$ on energy. 
    For large classicalons,  $r_* \gg L_*$,  the  dependence can be parameterized by a power-law, 
    $r_* \, \sim  \,  L_* (L_*m)^{\gamma}$, where $\gamma$ is a parameter, that defines how efficiently 
    the field is (self)sourced by the energy.  The quantization rule, now can be found out 
    from the requirement, that classicalon production cross section  for high $m$, goes 
    as a geometric cross section $r_*^2$.   Which, just like black holes implies the quantization 
    of mass in units of $r_*$, according to (\ref{rule1}). Translated for a mass we get, 
      \begin{equation}
      m =   N^{{1\over 1 \, + \, \gamma}} L_*^{-1} \, .
      \label{classquant}
      \end{equation}
      Our results  are fully understandable in the light of findings of ref\cite{gia-cesar-alex}, which 
      makes close  parallel between the generic classicalons  and black holes in terms of 
   field configurations composed out of  $N$ soft (wavelength $\sim r_*$)  bosons.  The quantization  rule (\ref{rule1}) is then just represents a requirement that the number of bosons is quantized.  
   
 %    Due to this, the quantization  rule can also  be translated  in terms of   the generalized   holographic  bound on information storage formulated in \cite{gia-cesar-alex}.   The role of the bound on information storage is clear from the fact that a generic classicalon  is a superposition  of   $N$ soft quanta of  wave-length $r_*$ and number given by (\ref{rule1}).    The classicalon quantization simply tells us that in large wave-length limit, the quanta are closer and closer to cold bose-condensate and the number of bosons in this condensate accounts for the log of possible number of information bits  stored in the configuration.    

\section{Level-Splitting?}

 We have proven the necessity of the discreteness of the black hole states  along the 
mass vertical.   A  level of a fixed mass $m$ is expected to have an internal degeneracy,  
which we shall refer to as the {\it horizontal}  degeneracy. 

 Such a degeneracy would mean that  a black hole of mass $m$ can be in a number of different 
 internal states which we can further label by index $a$, with all of them  having  a common mass  $m_a$.  When we discuss a production rate of  a black hole of a given mass $\Gamma(m)$,  all such internal degeneracies are already summed up. That is,  
 \begin{equation}
 \Gamma(m) \, = \, \sum_a  \Gamma(m_a) \, .   
 \label{splitting}
 \end{equation}
 The existence of a horizontal  degeneracy 
 is suggested by the Bekenstein's entropy counting arguments, and is usually estimated to be 
 $\sim \, e^{mr_*}$.   Since we showed that for black holes  the quantity $mr_*$ is a discrete number $N$,   our findings have some bearing on the above horizontal  degeneracy and suggest that 
 horizontal degeneracy is also  discrete and given by $e^{N}$. 
 
  Can one use the latter statement  against our proof, and suggest  that discreteness of the degeneracy could result into the level-splitting which could effectively  "wash-out" the discreteness 
  of the spectrum?   
  As we can easily see, this is not the case. 
  
    First, notice that there is absolutely no evidence that the levels can be split
    in a Poincare-invariant way.  Having a background field that violates asymptotic Pioncare 
    symmetry is beyond our interest.  An explicit example of such unremovable degeneracy is given by  string levels. So the first pint is that level-splitting doesn't follow from anything we know  about black hole physics.

     However,  even if we assume that  the degeneracy can be removed, this will not affect our proof. 
  Indeed,  such a level-splitting would populate the neighborhood  of  a  any mass $m$, with 
  exponentially large number of black hole states of distinct mass $m_a$. 
  This changes nothing in our previous reasoning, since all these new states continue to be 
 extremely sharp resonances.    What is extremely important to realize is,  that what count is 
 the relation between the total width $\Delta(m_a)$   of an individual resonance to its mass
 $m_a$, and {\it not}  a distance to a nearest level. 
  As long as the two levels $m_a$ and $m_b$ satisfy the condition
  \begin{equation}
    \Delta (m_a)  \, \ll \, m_a, ~~~~ \Delta(m_b)  \, \ll \, m_b \, , 
   \label{levelsplit}
  \end{equation}
 the distance between the levels $m_a - m_b$ is irrelevant. 
 This is obvious from the fact that when levels are not split $m_a = m_b$, we are back to the 
 old situation, when each state counts. 
 
   Now the remaining question is whether a densely populated spectrum can imitate the effect of the continuum.   To see that this cannot be the case, let us assume an extreme scenario in which 
   the split levels are uniformly spread.  In such a case the distance between the neighboring 
   levels instead of $\Delta m = r_*^{-1}$, now becomes  $\Delta m = r_* e^{-mr_*}$. 
    It seems that such closely spaced levels in any scattering experiment  will be indistinguishable 
    from continuum.   However,  this is just an illusion, in reality the resonances 
    that are produced by a geometric cross section $r_*^2$,  will continue to obey the quantization rule (\ref{rule1}), and be spaced 
 by $r_*^{-1}$. 

  To see that the resonances level  splitting must be set by $\Delta \bar{m} = r_*^{-1}$ and not by 
  $\Delta m = r_* e^{-mr_*}$, let us work in the approximation  in which 
  $\Delta m  \rightarrow  0$. In this limit we recover exactly the situation discussed in section 2.3. 
   The production rate of  each state of a definite mass $m$ now becomes infinitely suppressed, and in any scattering process instead we create a distribution, a resonance, characterized by a spectral function 
  $\rho_{\bar{m}}(m)$.   The point is that each of these distributions obeys the 
  condition of a sharp resonance, and thus counts in the total rate.  So the values of 
  $\bar{m}$ must be discrete, and moreover to reproduce the geometric cross-section
  the masses should obey the quantization rule (\ref{rule1}) in which $m$ is replaced by $\bar{m}$.

\subsection{Implications for LHC Black Hole Resonances}

    To finish this discussion, let us apply the above reasoning to a phenomenologicaly   
  most relevant case of black hole production  in the collision of  Standard Model  particles. 
  Let us consider a creation of a black hole resonance of mass $\bar{m}$, in a collision 
  process of some standard model particles
  and show that   the mass splitting of the observable sharp  resonances $\Delta \bar{m}$,  will be set by their production cross-section scale,  $1/\sqrt{\sigma}$,  even if  the  "true" mass eigenstate levels  are much closed spaced,   $\Delta m \, \ll \,  \Delta \bar{m}$.   
  
   Consider a collision process of some Standard Model particles in some in-state $|SM_{in}\rangle$, which results into a creation of a black hole resonance $|BH_{\bar{m}}\rangle$ and some 
   Standard Model particles in a final state $|SM_{out}\rangle$, 
   \begin{equation}
      SM_{in}  \, \rightarrow BH_{\bar{m}}  \, +  \, SM_{out} \, .
   \label{smtobh}
   \end{equation}
   Let us assume that  at the center of mass energy $\sqrt{s}$ the cross section of the above process is dominated by production  of a black hole resonance of mass $\bar{m} \sim \sqrt{s}$ 
 and is at least approximately given by a geometric area $\sigma \sim r_*^2(\bar{m})$
 according to (\ref{cross-section}).   A precise dependence of $r_*$ on $\bar{m}$ is unimportant, as long as $r_*$ is a dominant 
 length scale, growing with $\bar{m}$.  
  
   We should keep in mind that the state $|BH_{\bar{m}} \rangle$ we observe in this process 
   is a distribution of the "true" mass levels, $m$,  
   \begin{equation}
 |\bar{BH}_{\bar{m}}\rangle \,   =  \,  \sum_m \, \psi_{\bar{m}}(m) \,  | BH_m\rangle \, ,  
  \label{superposition1}
  \end{equation} 
 where the coefficients  $\psi_{\bar{m}}(m)$ define the relative weights by which an $m$-th black hole level  
 enters in a given  $\bar{m}$-th resonance,  and  $\rho_{\bar{m}}(m) \, \equiv \, |\psi_{\bar{m}}(m)|^2$ is a spectral density function.
  
   It is clear that the problem we are dealing with is a discretized version of the one already discussed in section 2, and in particular (\ref{superposition1}) is just a discretized version 
   of (\ref{superposition}).   However, since the level spacing is small, all the previous conclusions apply.  The only new thing is to adopt this conclusion to the constraint that the cross section 
  must be geometric, given by (\ref{cross-section}). 
    
     Let us now ask, how finely the resonance states $\bar{m}$ can be spaced? 
  Obviously, production of a nearest resonance  $\bar{m}'$  from the same initial state $|SM_{in}\rangle$ implies the change of the coefficients $\psi_{\bar{m}}(m)$, but  this is impossible without changing the Poincare-invariant  characteristics of the final state $|SM_{out}\rangle$.    In other words, $|SM_{out}' \rangle $ corresponding to a new resonance $\bar{m}'$  must be a {\it different } Poincare state.  For example, it has to include a different number and/or type of the Standard Model  particle species.   To be concrete,  we can imagine that a  creation of a 
  lighter  black hole of mass  $\bar{m}'   \, < \, \bar{m}$  in a same  two particle collision in 
  $|SM_{in} \rangle$ is accompanied by emission of more particle quanta in $|SM_{out}'\rangle$. 
  
  Obviously, such possible final states are discretized, and moreover  their number is not nearly enough to accommodate almost equal opportunities for the appearance of  black hole resonances 
  if they are more finely spaced than $\Delta \bar{m} \, \sim \, r_*(\bar{m})^{-1}$,  without being inconsistent with  the starting cross section.     
  
   We thus are reaching a surprisingly general prediction,  that the level spacing of 
observable sharp  resonances is restricted by their cross section scale, according 
to (\ref{crossrule}), 
 regardless of the underlying fine structure of mass eigenstates.

 \section{Conclusions}
 
 We are pointing out that there is a fundamental reason why the mass/energy spectrum 
 of any localized bound-state  cannot be continuous  and  must be quantized. 
  Physical reason is  that continuous spectrum of finite norm states  (stable or resonances) 
 would inevitably imply infinite production rate. 
 
 Mathematically,  discreteness of masses follows from the discreteness of the  finite-norm eigenfunctions of any hermitian operator.  

 And this constraint comes from first principles of quantum physics of any Poincare-invariant 
 asymptotic background and is independent  
  on particular short-distance properties of the state.   Black holes,  and classicalons in general,  are no exception from this rule.  
  
   Mass-quantization of  such states can be proven unambiguously from the first principles, 
  but the precise rule of quantization requires more input. 
  We have provided such an input, by linking the level spacing to the production cross-section.  
 This gives the rule (\ref{crossrule}).   
  Requiring that classicalon production cross section in high-energy scattering asymptotically approaches the geometric value given in terms of the area 
  $r_*^2$, for large $m$,  gives the  quantization rule 
(\ref{rule1}), which for particular case of Schwarzschild black holes implies that area is quantized 
  in the units of fundamental length $L_*$, in arbitrary number of dimensions.  
  
    Most important phenomenological implication of our results is, that micro  black holes and other classicalons that may be accessed by collider experiments   will come in form of quantum resonances, and 
    we can only distinguish them  from ordinary particles by probing higher and higher states. 
   Of course, for the lightest black holes, the quantization rule (\ref{rule1})  can be modified by order 
   one coefficients, but quantization will inevitably play a most important role there. 
    To uncover the precise form of the  quantization rule for lowest black hole resonances, we need more experimental input,  and we are looking forward to it.   
    
        \vspace{5mm}
\centerline{\bf Acknowledgments}

We thank  O. Kancheli and  G. 't Hooft for discussions. 
The work of G.D. was supported in part by Humboldt Foundation under Alexander von Humboldt Professorship,  by European Commission  under 
the ERC advanced grant 226371,  by TRR 33 \textquotedblleft The Dark
Universe\textquotedblright\  and  by the NSF grant PHY-0758032. 
The work of C.G. was supported in part by Grants: FPA 2009-07908, CPAN (CSD2007-00042) and HEPHACOS-S2009/ESP1473.
V.M. is   supported by TRR 33 \textquotedblleft The Dark
Universe\textquotedblright\ and the Cluster of Excellence EXC 153
\textquotedblleft Origin and Structure of the Universe\textquotedblright .

\end{document}